\documentclass[preprint,amssymb]{revtex4-1}%
\usepackage{amsmath,amssymb}
\usepackage{revsymb4-1}
\usepackage[cmyk]{xcolor}
\usepackage[%
	colorlinks=true,%
	linkcolor=blue,%
	citecolor=blue,%
	urlcolor=blue
	]{hyperref}%
\usepackage{graphicx}%
\usepackage{here}
\usepackage{dcolumn}%
\usepackage{bm}%
\usepackage{revsymb}
\usepackage{braket}
\usepackage{layouts} 
\usepackage{color}

\usepackage{txfonts}
\usepackage{comment}
\usepackage{braket}  

\def\equationautorefname~#1\null{\textrm{~(#1)\;}\null}
\def\figureautorefname~#1\null{Fig.~#1\null}

\begin{document}

\title{Spatial Search on Sierpinski Carpet Using Quantum Walk}

\author{Shu Tamegai, Shohei Watabe, and Tetsuro Nikuni}
\affiliation{Tokyo University of Science, 1-3 Kagurazaka, Shinjuku-ku, Tokyo, 162-9601, Japan}

\begin{abstract}
We investigate a quantum spatial search problem on a fractal lattice. A recent study for the Sierpinski gasket and tetrahedron made a conjecture that the dynamics of the search on a fractal lattice is determined by spectral dimension. We tackle this problem for the Sierpinski carpet, and our simulation result supports the conjecture. We also propose a scaling hypothesis of oracle calls for the quantum amplitude amplification. 
\end{abstract}

\maketitle

\begin{figure}[b]
\centering
	\includegraphics[width=7.4cm]{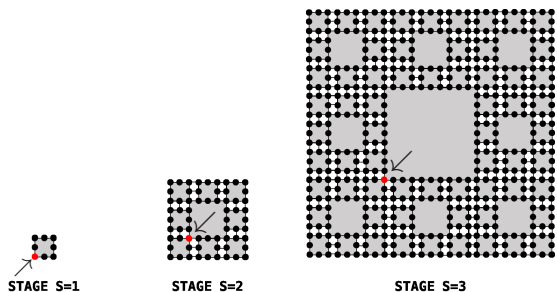}
	\caption{(Color online) Sierpinski carpets at various stages. The red vertex pointed by the arrow at each stage represents the marked vertex.}
	\label{fig:stage}
\end{figure}

The quantum algorithms for the spatial search problem have been intensively and extensively studied~\cite{PhysRevA67,PhysRevA82032331,Sci20999,PhysRevA82032330,PhysRevA86012332,SIAMPhiladelphia2005}. 
One expects that the superposition of states can efficiently find a marked object from a huge database. 
The quantum spatial search algorithm evolves the state by alternating the oracle and quantum walk operators. 
In order to solve this spatial search problem, the number of oracle calls should be optimized to concentrate the probability amplitude at the marked vertex. 
Understanding the scaling behavior of the optimal oracle calls is an important problem, which has been studied in a variety of regular lattice geometries~\cite{PhysRevA67,Sci20999,PhysRevA82032330,PhysRevA82032331}. 
The number of interval steps between the peaks of the marked vertex probability is given by $D N^{1/D}$ or $\pi \sqrt{N} /4$~\cite{PhysRevA82032330,PhysRevA86012332}, where $D$ is the dimension, and $N$ the number of sites. 

In fractal geometry, there are three characteristic dimensions---the Euclidean dimension $d_{\rm E}$, fractal (or the Hausdorff) dimension $d_{\rm f}$, and spectral (or the fracton) dimension $d_{\rm s}$---, so the question is which dimension determines the scaling behavior of the optimal number of oracle calls in a fractal lattice. 
Recently, an interesting conjecture for a fractal lattice was made by Patel and Raghunathan~\cite{PhysRevA86012332} that the scaling behavior of the spatial search is determined by the spectral dimension $d_{\rm s}$, and neither by the Euclidean dimension $d_{\rm E}$ nor by the fractal dimension $d_{\rm f}$. 
This conjecture was derived based on numerical studies for Sierpinski gasket ($d_{\rm E} = 2$) and Sierpinski tetrahedron ($d_{\rm E} = 3$). 
In this study, we investigate this conjecture for other fractal lattice geometry, the Sierpinski carpet ($d_{\rm E} = 2$), as in Fig.~\ref{fig:stage}. 
We show that our numerical simulation supports the conjecture made by Patel and Raphunathan~\cite{PhysRevA86012332}. 
We also propose the scaling hypothesis of the effective number of oracle calls for the quantum amplitude amplification.

In the quantum spatial search, we employ the flip-flop walk~\cite{SIAMPhiladelphia2005} as the quantum walk, 
where the state $\ket{\psi (t)}\equiv \sum_{x,l}a_{x, l} (t) \ket{\vec{x}}\otimes \ket{\hat{l}}$ is constructed in the Hilbert space $\mathcal{H}_{\rm search} \equiv \mathcal{H}_N\otimes \mathcal{H}_k$. 
Here, $\ket{\vec{x}} \in \mathcal{H}_N$ is associated with the position degree of freedom, and $\ket{\hat{l}} \in \mathcal{H}_k$ is associated with the link degree of freedom~\cite{PhysRevA86012332}. 
In the case of Sierpinski carpets, we take the dimension of $\mathcal{H}_{k}$ as $k=4$, for simplicity. 

\begin{figure}[tb]
	\centering
	\includegraphics[width=7.5cm]{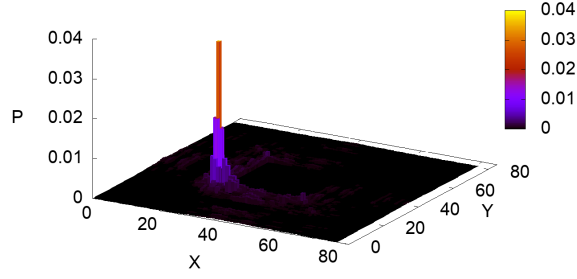}
	\caption{(Color online) Probability distribution for quantum spatial search on a Sierpinski carpet in the case when the amplitude distribution is concentrated toward the marked vertex. The data are for the stage $S=4$. }
	\label{fig:peak}
\end{figure}

\begin{figure}[tb]
	\centering
	\includegraphics[width=7.5cm]{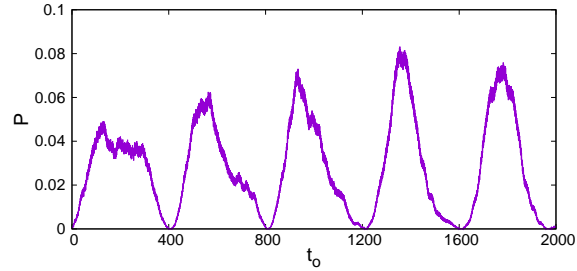}
	\caption{(Color online) Time evolution of the marked vertex probability $P(\vec{x} = \vec{0},  t)$ for the quantum spatial search on the Sierpinski carpet. The data are for the stage $S=4$. }
	\label{fig:iteration}
\end{figure}

The quantum spatial search algorithm is performed by the time evolution with alternately operating the oracle $R$ and the quantum walk $W$, i.e., $\ket{\psi (t)}= (W R)^{t}\ket{\psi (t= 0)}$. 
The oracle $R = R_{N} \otimes I_{k}$ with $R_{N} \equiv I_{N} -2\ket{\vec{0}}\bra{\vec{0}}$ gives the maximum contrast between a marked vertex at the origin $\vec{x} = \vec{0}$ and the rest. 
Here, $I_{N}$ and $I_{k}$ are the identity operators in $\mathcal{H}_{N}$ and $\mathcal{H}_{k}$, respectively. 
The quantum walk operator $W$ is composed of a Grover diffusion operator $G$ \cite{GroverLovK} and a shift operator $\mathcal{S}$, i.e., $W = \mathcal{S}G$. 
The Grover diffusion operator works as the inversion operator, given by~\cite{PhysRevA86012332} $a_{x, l}\xrightarrow{\;G\;} (2/k) \sum_{l'}a_{x, l'}-a_{x, l}$; in the case of the marked vertex, we do not operate the Grover diffusion operator, but the sign of the amplitude is flipped according to the oracle $R$~\cite{PhysRevA86012332}. 
The operator $\mathcal{S}$ shifts the amplitude along its link direction and reverses the link direction\cite{PhysRevA86012332}: $\ket{\vec{x}}\otimes \ket{\hat{l}}\xrightarrow{\;\mathcal{S}\;}\ket{\vec{x}+\hat{l}}\otimes \ket{-\hat{l}}$. 
If there is no vertex in the link direction $\hat{l}$, we take $\ket{\vec{x}}\otimes \ket{\hat{l}}\xrightarrow{\;\mathcal{S}\;}\ket{\vec{x}}\otimes \ket{\hat{l}}$. 
For the time evolution governed by the unitary operator $W = \mathcal{S}G$~\cite{PhysRevA86012332}, one may expect oscillation of the probability distribution as a function of the time step. 
In the present quantum spatial search, it is customary to choose the initial state to be a uniform superposition state: $\ket{\psi (  t = 0)} =(Nk)^{-1/2} \sum_{x,l} \ket{\vec{x}}\otimes \ket{\hat{l}}$.

\begin{figure}[tb]
	\centering
	\includegraphics[width=7.5cm]{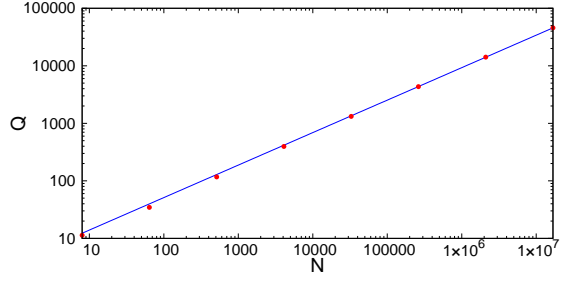}
	\caption{(Color online) Scaling of the optimal steps of the oracle calls $Q$ for the spatial search on the Sierpinski carpet. The linear fit is for the data from the stage $S=1$ to $S=8$. }
	\label{fig:fracton}
\end{figure}

\begin{figure}[t]
	\centering
	\includegraphics[width=7.5cm]{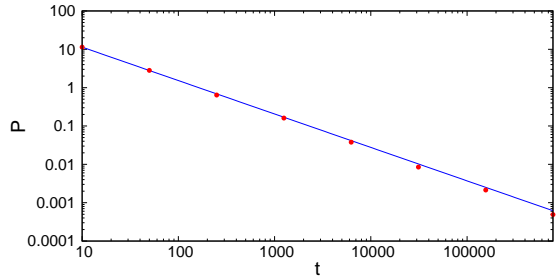}
	\caption{(Color online) Time evolution of return probability for a classical random-walker on a Sierpinski carpet. The linear fit is for the data of the stage $S=10$. }
	\label{fig:return}
\end{figure}

We numerically simulate the quantum spatial search algorithm on the Sierpinski carpet for stages $S=1$--$8$ ($N=8$--$16\,777\,216$). 
The probability distribution $P(\vec{x},  t) = \sum_{l} |a_{x,l} (t)|^{2}$ can be concentrated at the marked vertex in the Sierpinski carpet  (Fig.~\ref{fig:peak}). 
We reasonably find the oscillation of the probability at the marked vertex $P(\vec{x} = \vec{0},  t) = \sum_{l} |a_{0,l} (t)|^{2}$ (Fig.~\ref{fig:iteration}). Since the amplitude is small, one may need the quantum amplitude amplification. 
We identify the periodicity as the optimal time steps $Q$ to the oracle calls, which is extracted by using the Fourier transformation of the oscillatory data of the marked vertex probability $P(\vec{x} = \vec{0},  t)$. 
In our numerical simulation, we employed the number of steps to be approximately $1\,000\,000$. 
The size dependence of the optimal number of oracle calls $Q$ is shown in Fig.~\ref{fig:fracton}. 
Assuming $Q\propto N^b$, our fit gives the scaling behavior, $Q=3.79(4)N^{0.5647(6)}$, which gives the scaling exponent $b = 0.5647(6)$. 
We also study the mean value of the maximum marked vertex probability in the range of the period $Q$. Assuming $P \propto N^{-a}$, the fit of our data at stages $S=5$--$8$ gives the scaling behavior $P = 0.238(6)N^{-0.154(2)}$ with the scaling exponent $a = 0.154(2)$. 
In the following, we analyze this scaling exponent in connection with the fractal structure. 

In the Sierpinski carpet, the Euclidean dimension is given by $d_{\rm E} = 2$. 
The fractal dimension is defined as $d_{\rm f} \equiv \log{M(s)} / \log{s}$. 
Here, $M(s)$ is the number of the self-similar pieces, and $s$ the scale factor, where a line segment is broken into $s$-self-similar intervals with the same length. 
In the Sierpinski carpet, one has $d_{\rm f} = \ln 8/\ln3 = 1.892\,789 \cdots$. 
The spectral dimension $d_{\rm s}$ can be defined by the scaling behavior of the return probability of the classical random walk~\cite{RobertAMeyers}, given by $P_{\rm c} (\vec{x} = \vec{0}, t)\propto t^{-d_{\rm s}/2}$, where the classical random-walker starts from $\vec{x} = \vec{0}$ at $t=0$. 
Although the spectral dimension for the Sierpinski gasket is simply given by $d_{\rm s} = 2 \ln (d_{\rm E}+1)/ \ln (d_{\rm E}+3)$ \cite{JPhysLett43625,JPhysLett4413}, there is no explicit expression for the spectral dimension $d_{\rm s}$ of the Sierpinski carpet. 
We, therefore, determine the spectral dimension from the numerical simulation of the classical random walk on the Sierpinski carpet. 
From a simple scaling fit of our numerical simulation, we find $P_{\rm c}(\vec{0}, t)=84.7(9)t^{-0.871(4)}$ (Fig.~\ref{fig:return}), and thus, the spectral dimension of the Sierpinski carpet is given by $d_{\rm s}=1.742(8)$. 
Our result is consistent with the earlier analytic prediction for the range of the spectral dimension $1.673\,7\cdots \le d_{\rm s} \le 1.862\,0\cdots \;$ \cite{JMathVol51}.

We compare the spectral dimension with the scaling behavior of the optimal number of oracle calls $Q$. 
Using our numerical data, we find the approximate relation $b\simeq 1/d_{\rm s}=0.574(3)$, which supports the conjecture~\cite{PhysRevA86012332}. 
(In the case of the fractal dimension, we find $1/d_{\rm f}  = 0.528\,3 \cdots$.) 
Finally, we propose the scaling hypothesis for the effective number of oracle calls~\cite{PhysRevA86012332} $Q/\sqrt{P} \propto N^{b+a/2} \equiv N^c$ for the quantum amplitude amplification. Our hypothesis is given by the relation 
\begin{align}
c = d_{\rm s}/(d_{\rm E} - 1) + d_{\rm f} - s. 
\label{scaling_conjecture}
\end{align}
Using our data for the Sierpinski carpet ($s=3$), we find that the left hand side (l.h.s.) of Eq. (\ref{scaling_conjecture}) is $0.641(1)$ and the right hand side (r.h.s.) is $0.634(8)$, where the equality (\ref{scaling_conjecture}) holds within the standard error. 
By analyzing the data shown in Ref.~\cite{PhysRevA86012332} (with $s=2$), we find that the scaling hypothesis also holds well; for the Sierpinski gasket ($d_{\rm E} = 2$), the l.h.s. is $0.950(3)$ and the r.h.s. is $0.950\, 17 \cdots$, and for the Sierpinski tetrahedron ($d_{\rm E} = 3$), the l.h.s. is $0.773(4)$ and the r.h.s. is $0.773\, 70 \cdots$. Although no mathematically rigorous arguments exist, numerical results imply that the relevant scaling for the quantum spatial search may be given by $Q\propto N^{1/d_{\rm s}}$, and $Q/\sqrt{P} \propto N^{d_{\rm s}/(d_{\rm E} - 1) + d_{\rm f} - s}$.

In this study, we investigated the spatial search problem on a Sierpinski carpet using quantum walk. 
Our numerical simulation supports the recent conjecture that the scaling behavior of the quantum spatial search on a fractal lattice is determined by the spectral dimension, and not by the fractal dimension for the optimal oracle calls. 
We also proposed the scaling hypothesis of the effective number of oracle calls for the quantum amplitude amplification in a fractal lattice, which holds in the Sierpinski carpet, gasket, and tetrahedron. 
\begin{acknowledgments}
S.W. was supported by JSPS KAKENHI Grant No. JP16K17774.
T.N. was supported by JSPS KAKENHI Grant No. JP16K05504. 
\end{acknowledgments}
%
%


\end{document}